\documentclass[10pt,doublecolumn,twoside]{IEEEtran}
\IEEEoverridecommandlockouts
\usepackage{cite}
\usepackage{amsmath,amssymb,amsfonts}
\usepackage{algorithmic}
\usepackage{graphicx}
\usepackage{textcomp}
\usepackage{epstopdf}
\usepackage{float}
\usepackage{esint}
\usepackage{subfigure}
\usepackage{color}
\usepackage{mathrsfs}
\usepackage{algorithm}
\usepackage{algorithmic}
\usepackage{bm}
\usepackage{cases}
\usepackage{changes}

\newtheorem{theorem}{Theorem}
\newtheorem{lemma}{Lemma}

\def\BibTeX{{\rm B\kern-.05em{\sc i\kern-.025em b}\kern-.08em
    T\kern-.1667em\lower.7ex\hbox{E}\kern-.125emX}}

\makeatletter

\newcommand{\Rmnum}[1]{\expandafter\@slowromancap\romannumeral #1@}
\makeatother

\newcommand{\ls}[1]
    {\dimen0=\fontdimen6\the\font
     \lineskip=#1\dimen0
     \advance\lineskip.5\fontdimen5\the\font
     \advance\lineskip-\dimen0
     \lineskiplimit=.9\lineskip
     \baselineskip=\lineskip
     \advance\baselineskip\dimen0
     \normallineskip\lineskip
     \normallineskiplimit\lineskiplimit
     \normalbaselineskip\baselineskip
     \ignorespaces
    }

\graphicspath{{Code/}}
\begin{document}
\title{Age of Trust (AoT): A Continuous Verification Framework for Wireless Networks}

\author{Yuquan Xiao,~\IEEEmembership{Graduate Student Member,~IEEE}, Qinghe Du,~\IEEEmembership{Member,~IEEE}, Wenchi Cheng,~\IEEEmembership{Senior Member,~IEEE}, Panagiotis D. Diamantoulakis,~\IEEEmembership{Senior Member,~IEEE}, and George K. Karagiannidis,~\IEEEmembership{Fellow,~IEEE}

\thanks{Yuquan Xiao and Qinghe Du are with the School of Information and Communications Engineering, Xi'an Jiaotong University, Xi'an 710049, China. (e-mail:duqinghe@mail.xjtu.edu.cn)

Wenchi Cheng is with the State Key Laboratory of Integrated Services Networks, Xidian University, Xi'an 710071, China.

G. K. Karagiannidis and P. D. Diamantoulakis are with Department of Electrical and Computer Engineering, Aristotle University of Thessaloniki, Greece (e-mails: geokarag@auth.gr, padiaman@auth.gr).

}

}

\maketitle

\begin{abstract}
Zero trust is a new security vision for future networks that emphasises the philosophy of \emph{never trust and always verify}. However, there is a fundamental trade-off between the wireless transmission efficiency and the trust level, which is reflected by the verification interval and its adaptation strategy. More importantly, the mathematical framework to characterize the trust level of the adaptive verification strategy is still missing. Inspired by this background, we propose a concept called \emph{age of trust} (AoT) to capture the characteristics of the trust level degrading over time, with the definition of the time elapsed since the last trust verification of the target user. To evaluate the long-term trust level, the average AoT is used. We then investigate how to find a compromise between average AoT and wireless transmission efficiency with limited resources. In particular, we address the bi-objective optimization (BOO) problem between average AoT and throughput over a single link, where the identity of the receiver is constantly verified, and we devise a periodic verification scheme and a $Q$-learning-based scheme for constant and random service processes, respectively. We also tackle the BOO problem in a multiple random access scenario, where a trust-enhanced frame-slotted ALOHA is designed. Finally, the numerical results show that our proposals can achieve a fair compromise between trust level and wireless transmission efficiency, and thus have a wide application prospect in various zero-trust architectures.
\end{abstract}


\section{Introduction}
As the cornerstone of wireless networks, a variety of security provisioning paradigms have been developed and evolved over time~\cite{Ji2022Survey,Zou2016A,Yu2024Security,Liu2020Energy}, some of which have been ingeniously embedded in various communication standards such as WiFi, NearLink, 5G, etc., to secure data transmissions of all kinds, including enterprise services and personal private information among our environment. Traditional security provisioning paradigms advocate an entry-point identity verification principle, where once a user is authorised to enter a network, they are always trusted throughout the network's coverage until they are checked out. However, there is a potential risk that the authorised users could be hijacked by malicious attackers, resulting in information leakage and illegal occupation of resources~\cite{Liu2024Similarity,Lu2022Safe}, and leading to destructive consequences.

Zero trust is an emerging vision of security provisioning that is widely recognised as a replacement for the conventional entry point verification mechanism that suffers from the growing trust crisis after the one-time initial verification~\cite{zerotrust2010}. It emphasises never trust and always verify, and has significant potential to improve the resilience of the next generation of wireless communication networks~\cite{Ridhawi2023Zero,Xiao2024Space,Fei2023Air}. Zero trust networks assume that everything is always untrustworthy, even if you have been inside the network. In particular, zero trust seeks to minimise the attack surface through three approaches: least privilege access, micro-segmentation and continuous trust validation~\cite{Tsai2024Strategy}. Least privilege access encourages the source to give the user the minimum level of privileges needed to perform a given task. The second is to divide the network into one and one small segment, where each user only has the privilege to access the segments for which they are authorised. This reduces the attack surface in the space domain. Continuous trust verification aims to minimise the attack surface in the time domain by proposing that data sources constantly verify the identity of accessing users instead of the traditional one-time verification principle.

\subsection{Related Work}
\label{sec:state_of_art}
Next, we summarize the existing literature in the area of zero trust, with a particular focus on continuous trust verification.

The concept of zero trust with the philosophy of trusting but verifying was introduced by Forrester Research in 2010 to address potential threats~\cite{zerotrust2010}, such as internal attacks and eavesdropping, to the use of traditional trust verification principles. In the least-privilege access domain, the authors of~\cite{Poirrier2023An} proposed an enforced verification method for every access request, even if the target is a federated partner in the Internet of Military Things. In the area of microsegmentation, the authors of~\cite{Mujib2020Performance} validated that the use of such a measure can significantly improve the security performance of data center networks at a low time cost. In the area of continuous verification, the authors of~\cite{Yu2019Continuous} proposed a channel state information-based continuous verification scheme for status update systems, where a round of verification takes only 1.5 ms. The authors of\cite{Xu2024Distributed} discussed the problems of significant delay introduced by data verification in zero-trust architectures over large networks. There is also a lot of work being done to achieve zero trust in a comprehensive way. The authors of~\cite{Liu2024Zero} presented various candidate solutions to achieve zero trust for 6G networks in terms of identity authentication, access control, and confidential transmission, and envisioned the advanced technologies such as blockchain, generative AI, and federated learning~\cite{Zhan2024Cloud,Tang2022Secure,Hou2024Efficient}, to facilitate it. In \cite{Chen2021A}, a zero-trust awareness and protection system for remote healthcare was presented, aiming to evaluate trust based on subject, object, environment, and behavior. The author of~\cite{Hussain2024Federated} proposed the use of federated learning in artificial intelligence to develop zero-trust architectures, where privacy concerns can be inherently addressed through federated learning.

Continuous trust verification is essential in zero-trust architectures, with significant research focused on identifying users and assessing trust levels. Most of the existing work focuses on what means should be used to verify identity and assess trust at each verification. It has been recognized that classical password verification is obsolete~\cite{Gong2020Secure,Wang2018Security}. Instead, user and device biometrics, such as hardware fingerprinting, are becoming the mainstream of identity representation. In particular, the authors of~\cite{Ceccarelli2012Improving} proposed a trust protocol to adjust authorization timeouts based on the quality of the biometric as well as the trust in the user's activity. In~\cite{Xie2024Industrial}, the conventional neural networks were used to model and dynamically analyze the user's behavior to evaluate the trust value.
The authors of~\cite{Chahoud2023Towards} studied the client trust problems for federated learning, where the trust factor is continuously updated after the client has unsuccessfully completed the assigned tasks. In~\cite{XU2022365Non}, the authors proposed a computation-efficient verification scheme for federated learning framework where the communication resources are limited. Reference~\cite{Lins2016Dynamic} suggested that dynamic certification for cloud services should be developed to perform continuous trust verification, including constant data collection, transmission, and analysis. The authors of~\cite{Elmaghbub2024Domain} defined a device authentication layer within the zero-trust architecture, which extracts the hardware fingerprint using deep learning methods to verify the identity of devices. We note that the issue of decaying trust over time has not yet been explored, which represents a key breakthrough of this paper. Finally, incorporating soft-level measures into verification procedures is sometimes necessary ~\cite{brezolin2023method,kang2020programmable}. This approach allows for timely monitoring of user behavior to analyze and predict whether a user within the network might become a malicious node, potentially compromising the network's security and reliability~\cite{de2023Applying}.

\begin{figure*}
  \centering
  \includegraphics[scale=0.65]{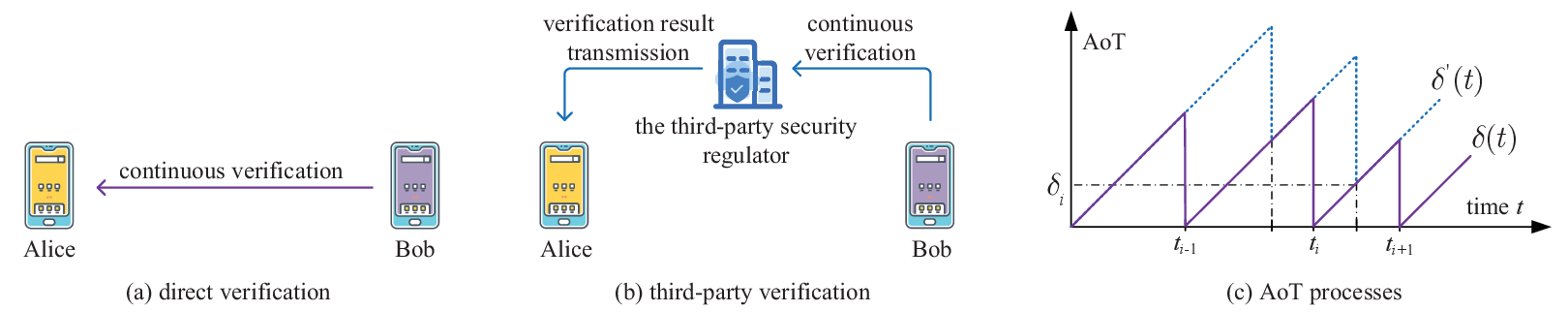}
  \caption{Two typical verification architectures in zero trust networks and the corresponding AoT processes.}\label{fig:AoT_process}
\end{figure*}

\subsection{Motivation and Contribution}
Continuous trust verification over wireless networks often needs to consume a certain amount of resources such as time, spectrum, energy, etc., for the interaction of verification information, which can sometimes be negligible in the traditional one-time entry point verification principle. In continuous trust verification, once the consumed resources reach a certain level, other conventional network performance, such as transmission efficiency, may be affected due to the reduction of available resources.  This is because the more frequent the verification, the more resources are consumed.  Thus, it is very challenging to achieve both zero trust and efficient use of available resources, which limits the level of security that can be achieved in several applications, such as the Internet of Things.  Sometimes, we implement zero trust by verifying the identity of every interaction, which could result in huge costs and lead to the transmission of low-efficiency regular services.  To this end, it is critical to find a trade-off between the level of trust reflected by the verification interval and its adaptation strategy, and other network performances such as transmission efficiency. If the security requirements are strict, we can devote more resources to performing verification procedures. If the security requirements are loose, we can devote more resources to improving other conventional network performance.

First, there is still no mathematical framework to characterize the trust level of the adaptive verification strategy. The longer the time since the last trust verification, the more severe the current trust crisis. Motivated by this time-cumulative nature, we would like to draw on the concept of Age of Information (AoI)~\cite{Kaul2012Real}, which has been widely used in status update systems to evaluate the freshness of status information~\cite{Huang2022Age,Xiao2024Adaptive,Sun2021AoI}. Therefore, we propose a concept called the age of trust to characterize the level of trust that degrades over time. Then, we investigate how to achieve the trade-off between AoT and other network performance. Specifically, the main contributions of this paper can be summarized as follows:
\begin{itemize}
  \item We propose the concept of \emph{age of trust} (AoT), which is defined as the amount of time that has elapsed since the target user's trust was last verified. The more frequent the verification, the smaller the AoT will be. The average AoT is further defined to assess the long-term confidence level. It is worth noting that some customised adjustments to the AoT can be made to meet the requirements of different scenarios.
  \item We study the bi-objective optimization (BOO) problem between average AoT and throughput over a single link with arbitrary service processes, and devise a periodic verification scheme and a $Q$-learning-based scheme for constant and random service processes, respectively. Heuristically, an improved periodic verification scheme is also designed for the random service process.
  \item We also investigate the BOO problem for performing continuous trust verification in a multiple random access scenario, where a trust-enhanced frame slotted ALOHA is designed with two types of time slots, i.e., the trust-enhanced time slot and the standard time slot. Unlike the standard time slot, the trust-enhanced time slot includes not only the regular data transmission phase but also the trust verification phase. In particular, we can control the percentage of trust-enhanced time slot to achieve the compromise between average AoT and throughput.
  \item Simulation results confirm that our proposals can achieve a good compromise between average AoT and throughput, and surprisingly show that our proposed improved periodic verification scheme can achieve the same performance as the proposed $Q$-learning-based scheme. It is also shown that the overhead of trust verification has a significant impact on the overall performance in a random access scenario.
\end{itemize}

The rest of this paper is organized as follows.  Section~\ref{sec:AoT} provides the definition of AoT as well as its characteristics. Section~\ref{sec:AoT_app_one} showcases the application of AoT for continuous trust verification over a single link with arbitrary service processes. Section~\ref{sec:AoT_app_two} investigates the application of AoT for multiple random access scenario. Section~\ref{sec:numerical_results} presents the effectiveness of our proposals. Finally, this paper is concluded with Section~\ref{sec:conclusions}.

\section{Age of Trust (AoT)}
\label{sec:AoT}
Secure data transmission is crucial in modern wireless networks, which would be vulnerable to various potential threats and attacks~\cite{Tinshu2023A}. Zero trust is expected to provide solid protection for wireless networks through three fundamental principles, namely least privilege access, micro-segmentation and continuous trust verification. However, continuous trust verification often requires a certain level of resources in wireless networks, which could weaken other network performance, such as transmission efficiency. It would not be practical to verify the identity of target users and assess the trust level at the beginning of each interaction. Therefore, there is a trade-off between the level of trust verification and other conventional network features.

However, the mathematical framework for characterising the trust level of adaptive verification strategies remains an open issue. It is recognized that the longer the time since the last trust verification, the higher the potential risk. Motivated by this observation, we introduce a concept termed Age of Trust (AoT), which is defined as \emph{the time that has elapsed since the target user was last verified}. The definition of AoT is inspired by the concept of AoI that is widely used to evaluate the freshness of information in various status update systems~\cite{Kaul2012Real,Lin2021Cooperative}. While AoI was originally developed to measure the freshness of status updates in data streams, we identify a fundamental parallelism between Information Freshness and Trust Validity in zero trust networks. In traditional perimeter-based security, trust is binary and static. However, in zero trust architectures, the environment is assumed to be hostile and dynamic. We introduce the AoT not merely as a direct application of AoI, but to quantify the accumulation of security uncertainty over time. This is based on the following rationale: 1) In a dynamic wireless environment, a user's security status is not static. Once verified, the certainty of the user's legitimacy is high. However, as time elapses without re-verification, the probability of adverse events, e.g., session hijacking, device compromise, or anomalous behavior, accumulates. 2) The linear growth of AoT serves as a proxy for the exposure time or the attack window. The longer the system goes without verification, i.e., the larger the age, the wider the window of opportunity for an attacker to exploit the link. Moreover, while they share a mathematical structure, they differ fundamentally in their physical semantics and trade-offs: 1) AoI measures the freshness of the data payload, such as temperature and humidity data packets in smarthome, whereas AoT measures the validity of the node's identity, e.g., via hardware fingerprint, for security consideration. 2)~AoT verification competes for resources with regular data transmission services, creating a unique trade-off between Security and Throughput, which is rarely considered in AoI literature.

Then, as shown in Fig.~\ref{fig:AoT_process}, we consider two typical verification architectures, namely, direct verification and third-party verification, and characterize the corresponding AoT processes. For the direct verification, as presented in Fig.~\ref{fig:AoT_process}(a), Alice verifies the identity of Bob locally, e.g., via hardware radio-frequency (RF) fingerprint. In such cases, the verification result is available immediately. Following the definition of AoT, the corresponding AoT process is characterized as $\delta(t)$ in Fig.~\ref{fig:AoT_process}(c), where $t_i$ denotes the time of the $i$th verification. Specifically, once Bob is verified, $\delta(t)$ is updated to zero. Between the interval of two adjacent verifications, $\delta(t)$ increases linearly with time. Here, we assume an ideal verification mechanism such that a successful verification completely refreshes the trust status of the target user, i.e., the AoT is reset to zero. This assumption allows us to establish a fundamental analytical framework for characterizing the relationship between verification frequency, trust freshness, and communication efficiency. In practical zero-trust systems, verification mechanisms may suffer from imperfect detection accuracy due to measurement noise, channel uncertainty, or limited authentication capability. In such cases, the AoT reset value can be generalized to a non-zero residual age \(\delta_{\min}\), whose relationship with verification accuracy and attack behaviors requires further investigation. For the third-party verification, as shown in Fig.~\ref{fig:AoT_process}(b), the verification is performed by a third-party verifier, which then transmits the verification result to Alice, thereby incurring a verification latency $\delta_i$ at the side of Alice. Accordingly, the AoT process at the third-party verifier is shown as $\delta(t)$ in Fig.~\ref{fig:AoT_process}(c), whereas the AoT process at Alice is shown as $\delta'(t)$, where the age is reset to $\delta_i$ at the $i$th verification. In this paper, we focus on the direct verification architecture. This basic scenario allows us to concrete on characterizing the fundamental trade-off between trust freshness and transmission efficiency. It is worth mentioning that a wide range of composition functions or slight adjustments in terms of AoT can be made to fulfill the requirements of various scenarios. For example, to characterize the mapping between AoT and the actual security risk or expected security loss, we can define appropriate risk mapping functions based on AoT. More specifically, consider a representative attack model where malicious events follow a Poisson process with arrival rate \(\lambda_a\). Given an AoT value \(\delta(t)\), the probability of at least one malicious event occurring during the unverified period is given by \(1-\exp(-\lambda_a \delta(t))\), which monotonically increases with AoT, demonstrating that AoT captures the temporal accumulation of security exposure.

To evaluate the long-term AoT performance, the average AoT, denoted by $\Delta$, is defined as
\begin{align}
\Delta = \lim_{T\rightarrow\infty}\frac{1}{T}\int_{0}^{T}\delta(t)dt,
\end{align}
where $T$ represents the duration of interests. Note that the average AoT will approach to the infinity if the conventional one-time verification principle is used.

As discussed earlier, the pursuit of ultimate zero trust could lead to a severe degradation of other network services due to the enormous resource cost of frequent trust verification. Also, different services have different security requirements, ranging from loose to strict. Sometimes we do not need to achieve strict zero trust. Therefore, the better choice is to find a fair compromise between AoT and other network services. To this end, we show below how the concept of AoT can be used to guide system design while validating its effectiveness.

\section{Continuous Trust Verification for Single-Link Transmissions}
\label{sec:AoT_app_one}
In this section, we investigate the problem of how to achieve the trade-off between average AoT and throughput for single-link transmissions. First, the system model is elaborated on. Then, the optimal trust verification schemes for constant service process and random service process are exploited.

\subsection{System Model}
Consider a wireless transceiver system where time is discretised as one and one equal-length time slot and the associated index is denoted by $t$ with $t = 1,2,3,...$ . The service rate of the transmitter at the $t$th time slot is denoted by $\mu(t)$. As the duration of one time slot is short, we assume that it is constant within a time slot, but varies between time slots. The identity of the receiver should be constantly verified throughout the data transmission process to combat the man-in-the-middle (MITM) attack. We take the hardware fingerprinting measure to verify the identity of the receiver~\cite{Yu2019Continuous}. In particular, a verification procedure will consume a time slot through which the receiver transmits a specific data frame to the transmitter for hardware fingerprinting extraction. When the verification procedure is in progress, the wireless channel is occupied by the receiver and the regular data service from the transmitter to the receiver is suspended. We define the following indicator function, denoted by $\mathcal{I}(t)$, to indicate whether the identity is being verified at time slot $t$, i.e,
\begin{equation}
\mathcal{I}(t)=\left\{ {\begin{array}{*{20}{l}}
1,&\mbox{if verifying the identity,}\\
0,&\mbox{otherwise,}
\end{array}} \right.
\end{equation}
where $\mathcal{I}(t) = 1$ indicates verifying the identity of the receiver at time slot $t$, and $\mathcal{I}(t) = 0$ points carrying the regular data service. Following a such configuration, the AoT at the $t$th time slot, denoted by $\delta(t)$, can be presented as follows:
\begin{equation}
\delta(t)=\left\{{\begin{array}{*{20}{l}}
0,&\mbox{if $\mathcal{I}(t) = 1$,}\\
\delta(t-1) + 1,&\mbox{if $\mathcal{I}(t) = 0$,}
\end{array}} \right.
\end{equation}
which shows once the identity of the receiver is being verified, the AoT is reset to zero, noting that we absolutely trust the receiver if its identity is verified, and otherwise, the AoT increases by one. Then, the average AoT $\Delta$ can be given as follows:
\begin{align}\label{eq:cd_average_aot}
\Delta = \lim_{T\rightarrow\infty}\frac{1}{T}\sum_{t = 1}^{T}\delta(t).
\end{align}
Moreover, the throughput, denoted by $\bar{\mu}$, can be formulated as a function of the indictor as well as the instantaneous service rate, $\mu(t)$, that is,
\begin{align}\label{eq:cd_average_rate}
\bar{\mu} = \lim_{T\rightarrow\infty}\frac{1}{T}\sum_{t = 1}^{T}\left[1-\mathcal{I}(t)\right]\mu(t),
\end{align}
which indicates the service rate will be zero if verifying the identity. Our objective is to find out the verification scheme which can maximize the throughput meanwhile minimizing the average AoT. This is a bi-objective optimization (BOO) problem, which is denoted by \textbf{P1} and formulated as follows:
\begin{align}
\textbf{P1}:
\max_{\mathcal{I}(t), \forall t} \,\,\, &(\bar{\mu},-\Delta),
\end{align}
where a negative sign is added at front of the average AoT for minimization. The weighted sum method is often used to solve BOO problems in light of its simple-deploying characteristics~\cite{marler2010weighted}. It should be noted that the weighted-sum method generally obtains supported Pareto-optimal solutions. Due to the integer-valued decision variables in our problems, the complete Pareto frontier may not be recovered if the underlying trade-off region is non-convex. In this paper, we adopt the weighted-sum method to characterize the achievable trade-off between AoT and throughput, which provides a computationally efficient approach for practical system design. With this method, \textbf{P1} can be converted to \textbf{P2} along with substituting Eqs. (\ref{eq:cd_average_aot}) and (\ref{eq:cd_average_rate}) into the objective as follows:
\begin{align}
\textbf{P2}:
\max_{\mathcal{I}(t),\forall t} \,\,\, \lim_{T\rightarrow\infty}\frac{1}{T}\sum_{t = 1}^{T}\left\{\left[1-\mathcal{I}(t)\right]\mu(t) - \alpha\delta(t)\right\},
\end{align}
where $\alpha\geq 0$ is the weighted factor associated with the average AoT. Since the AoT process is a Markov process, \textbf{P2} is a Markov decision process (MDP) problem. Next, we address \textbf{P2} for the cases of constant service process and random service process, respectively.

\subsection{Constant Service Process}
\begin{figure}
  \centering
  \includegraphics[scale=0.67]{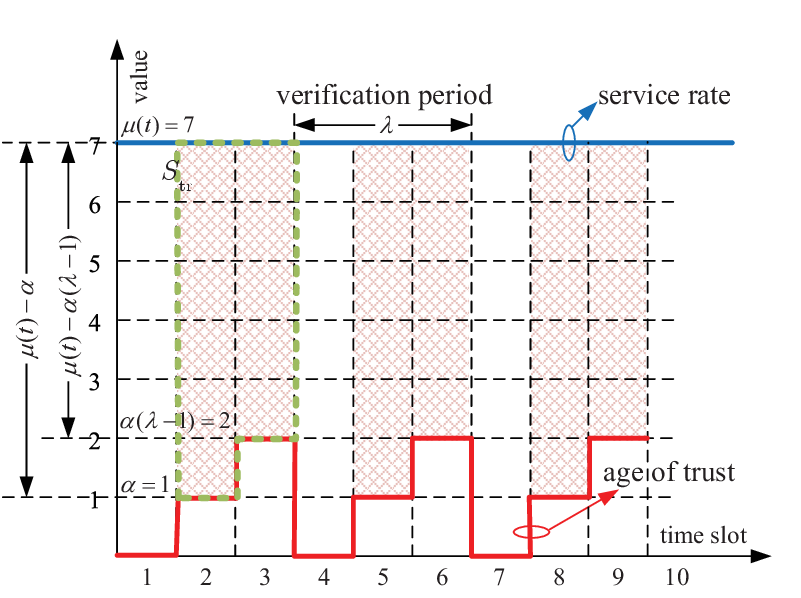}
  \caption{Continuous trust verification under a constant service process, where $\mu(t) = 7$ packets/slot, $\lambda = 3$ slots, and $\alpha = 1$.}\label{fig:AoT_constant}
\end{figure}
To facilitate the understanding of \textbf{P2} for the constant service process, we visualise the service process and the AoT process in Fig.~\ref{fig:AoT_constant}, where the service rate is set to 7 and the weighted factor is set to 1, for example. Since \textbf{P2} is an MDP problem and the corresponding state is the instantaneous AoT, which is deterministically changed, the optimal trust verification scheme must be periodic. We denote the verification period by $\lambda$. As shown in the figure, the AoT increases linearly within the verification period and is updated to zero when the verification is performed at the beginning of the next period. The shadow part at each time slot in the figure corresponds to the value of $\left[1-\mathcal{I}(t)\right]\mu(t) - \alpha\delta(t)$, which can be considered as the instantaneous reward associated with this MDP problem. In particular, if $\mathcal{I}(t) = 1$, i.e. we decide to verify the identity in this time slot, the instantaneous reward is zero. \textbf{P2} aims to find the verification scheme that can maximise the long-term reward. This is equivalent to finding the verification period that can implicitly maximise the shadow area over all time. In the following, we use a graphical method to find this optimal verification period. The details are given by Theorem~\ref{th:opt_ver_period_c}.
\begin{theorem}\label{th:opt_ver_period_c}
The optimal verification period, denoted by $\lambda^*$, for the constant service process is determined by
\begin{align}\label{eq:opt_ver_period}
\lambda^* = \left\lceil \sqrt{\frac{2\mu(t)}{\alpha}} \right\rceil\mbox{~or~}\left\lfloor \sqrt{\frac{2\mu(t)}{\alpha}} \right\rfloor,
\end{align}
where $\lceil\cdot\rceil$ and $\lfloor\cdot\rfloor$ denote the ceil and floor operators, respectively. Further, the one yielding a higher objective value for Problem~\textbf{P2} is selected.

\end{theorem}
\begin{IEEEproof}
The objective of \textbf{P2} corresponds to the time-average of shadowing area in Fig.~\ref{fig:AoT_constant}. Further, the shadowing area is composed of one and one trapezoid-shaped block, which is denoted as $S_{\rm tr}$ and equal to
\begin{align}
S_{\rm tr} = \frac{1}{2} (\mu(t)-\alpha +\mu(t) - \alpha(\lambda - 1))(\lambda-1).
\end{align}
The number of trapezoid-shaped blocks is $\frac{T}{\lambda}$ within the duration of interests $T$. Thus, the objective of \textbf{P2}, denoted by $f(\lambda)$, is equal to
\begin{align}
f(\lambda) =\lim_{T\rightarrow\infty}\frac{1}{T} \frac{T}{\lambda} S_{\rm tr}=\frac{(2\mu(t) -\alpha \lambda)(\lambda-1)}{2\lambda}.
\end{align}
Taking the derivative of $f(\lambda)$ with respect to $\lambda$ and letting the result equal to zero, we have
\begin{align}
\frac{df(\lambda)}{d\lambda} = 0 \Rightarrow \lambda = \sqrt{\frac{2\mu(t)}{\alpha}}.
\end{align}
Since $\lambda$ should be an integer, we have Eq.~(\ref{eq:opt_ver_period}).
\end{IEEEproof}

Theorem~\ref{th:opt_ver_period_c} indicates that the higher the service rate, the longer the verification period. This is because at higher service rates, the throughput will be greatly reduced if we perform the verification frequently. It also shows intuitively that the larger the weighting factor is, the shorter the verification period will be.

\subsection{Random Service Process}
In the following, we exploit the optimal verification scheme for the case of random service process. There are two subcases: 1) only the average service rate is known; 2) the instantaneous service rate is observable. The rationale for considering the first subcase is that sometimes the service rate is fast time-varying, so we cannot capture the instantaneous service rate in a timely manner, leading us to decide whether to verify the identification at the next time slot using only the information of the average service rate. On the other hand, for the slowly time-varying service rate, we can scarify a small amount of time at the beginning of each time slot to acquire the value of the current service rate, being like as the channel state information acquisition. We then decide whether to verify the identity at the beginning of the current time slot with not only the AoT but also the available instantaneous service rate, which corresponds to the second subcase.

\subsubsection{Only the Average Service Rate is Known}
When only the average service rate is available, the optimal verification scheme is the same as for the constant service process. The reason is that the instantaneous reward for this subcase is $\left[1-\mathcal{I}(t)\right]\mathbb{E}[\mu(t)] - \alpha\delta(t)$, where $\mathbb{E}[\cdot]$ is the mathematical expectation. It is similar to the constant service process, where the instantaneous reward varies purely with the instantaneous AoT and is independent of the instantaneous service rate, because we cannot capture it in time. The optimal verification scheme for this case is described in Theorem~\ref{th:opt_ver_period_r1}.
\begin{theorem}\label{th:opt_ver_period_r1}
The optimal verification period, denoted by $\widetilde{\lambda}^*$, for the random service process, where only the average service rate is known, is determined by
\begin{align}\label{eq:opt_ver_period_r1}
\widetilde{\lambda}^* = \left\lceil \sqrt{\frac{2\mathbb{E}[\mu(t)]}{\alpha}} \right\rceil\mbox{~or~}\left\lfloor \sqrt{\frac{2\mathbb{E}[\mu(t)]}{\alpha}} \right\rfloor.
\end{align}
Further, the one corresponding to a higher objective value of Problem~\textbf{P2} is selected.
\end{theorem}
\begin{IEEEproof}
\begin{figure}
  \centering
  \includegraphics[scale=0.55]{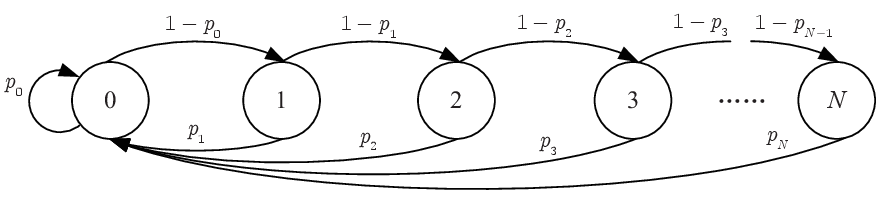}
  \caption{The Markov chain of AoT, where the service rate is equivalent to zero when we occupy the current time slot to conduct the trust verification.}\label{fig:Markov_AoT_process}
\end{figure}
We intend to obtain the optimal state transition probability for the Markov chain of AoT, i.e., maximizing the objective of \textbf{P2}. Specifically, the maximum AoT is denoted as $N$. Then, the state transition diagram for AoT is illustrated in Fig.~\ref{fig:Markov_AoT_process}. There are $N+1$ states for AoT, i.e., from 0 to $N$. At the $n$th state, the next state is either $n + 1$ or $0$, which is determined by whether we conduct the trust verification or not. If we verify the trust level, the AoT is reset to zero. Otherwise, the AoT increases by one. We denote by $p_n$ the probability of conducting the verification at the $n$th state, and $\pi_n$ the stationary probability of the state $n$. Then, based on the state transition diagram in Fig.~\ref{fig:Markov_AoT_process}, the state transition equation can be constructed as follows:
\begin{subequations}
\begin{numcases}{\hspace{-0.7cm}}
(1-p_0)\pi_0 = p_1\pi_1+...+p_N\pi_N \\
\pi_1 = (1-p_0)\pi_0 \\
\pi_2 = (1-p_1)\pi_1 \\
\hspace{1cm}\vdots \\
\pi_N = (1-p_{N-1})\pi_{N-1}\\
\sum_{n = 0}^{N}\pi_n = 1
\end{numcases}
\end{subequations}
Solving this equation, we have
\begin{subequations}\label{eq:sta_pi}
\begin{numcases}{\hspace{-0.7cm}}
\pi_0 = \frac{1}{1 + \sum_{n = 1}^{N}\prod_{i = 0}^{n-1}(1-p_i)},\\
\pi_n = \frac{\prod_{i = 0}^{n-1}(1-p_i)}{1 + \sum_{n = 1}^{N}\prod_{i = 0}^{n-1}(1-p_i)}, n = 1, 2, ..., N,
\end{numcases}
\end{subequations}
which evidently indicates
\begin{align}
\pi_0 \geq \pi_1 \geq \pi_2 \geq ... \geq \pi_N \geq 0.
\end{align}
Moreover, by using the notation of the stationary distribution of AoT, i.e., $\pi_0$, $\pi_1$, ..., and $\pi_N$, the objective of \textbf{P2}, denoted by $g(\pi_n)$, can be reorganized from the perspective of statistical average as follows:
\begin{align}
g(\pi_n) &= \pi_0\cdot 0 + \sum_{n = 1}^{N} \pi_n(\mathbb{E}[\mu(t)]-\alpha n) \\
&= (1-\pi_0) \mathbb{E}[\mu(t)] - \alpha \sum_{n = 1}^{N} n\pi_n,
\end{align}
which is a linear combination of $\pi_0$, $\pi_1$, ..., and $\pi_N$. Therefore, we can convert \textbf{P2} to a linear programming problem as follows:
\begin{subequations}\label{eq:lpp}
\begin{align}
\max_{\pi_n,\forall n} \,\,\, &g(\pi_n),\\
{\rm s.t.}\,\,\, & \pi_0 \geq \pi_1 \geq \pi_2 \geq ... \geq \pi_N \geq 0, \label{eq:lp_ex_1}\\
& \sum_{n = 0}^{N}\pi_n = 1.\label{eq:lp_ex_2}
\end{align}
\end{subequations}
Since the feasible region of \eqref{eq:lpp} is a nonempty polytope, there exists at least one optimal solution attained at an extreme point of the feasible region. Therefore, we only need to evaluate the objective function at the extreme points characterized in Lemma~\ref{lm:1}.

Substituting the given extreme points in Lemma 1 into $g(\pi_n)$, we obtain
\begin{align}
g(\pi_n) = \left(1-\frac{1}{1+n}\right)\mathbb{E}[\mu(t)] - \frac{\alpha n}{2}.
\end{align}
Taking the derivative of $g(\pi_n)$ with respect to $n$, we have
\begin{align}
\frac{dg(\pi_n)}{dn} = \frac{1}{(1 + n)^2}\mathbb{E}[\mu(t)] - \frac{\alpha}{2}.
\end{align}
By setting $\frac{dg(\pi_n)}{dn}$ to zero, we derive the optimal $n$ as follows:
\begin{align}\label{eq:opt_AoT_max}
\frac{dg(\pi_n)}{dn} = 0 \Rightarrow 1 + n = \sqrt{\frac{2\mathbb{E}[\mu(t)]}{\alpha}}.
\end{align}
Combining Eqs.~(\ref{eq:sta_pi}) and the result in Lemma~\ref{lm:1}, we have
\begin{align}\label{eq:opt_stp}
p_0 = p_1 = p_{n - 1} = 0,~\mbox{and}~p_n = ... = p_N = 1.
\end{align}
Eq. (\ref{eq:opt_stp}) indicates that the trust verification is performed if and only if the AoT is $n$. Therefore, there exists a verification period, and the corresponding optimal value is determined by Eq.~(\ref{eq:opt_AoT_max}), and described in Theorem~\ref{th:opt_ver_period_r1}.
\end{IEEEproof}

\begin{lemma}\label{lm:1}
The extreme points of the feasible region defined in \eqref{eq:lp_ex_1}-\eqref{eq:lp_ex_2} are given by
\begin{align}
\pi_0=\pi_1=\cdots=\pi_n=\frac{1}{n+1}, \label{eq:f_lp_ex_1}
\end{align}
and
\begin{align}
\pi_{n+1}=\cdots=\pi_N = 0, \label{eq:f_lp_ex_2}
\end{align}
where \(n=0,1,\ldots,N\).
\end{lemma}
\begin{IEEEproof}
To characterize the extreme points of the feasible region, we first introduce the following transformation:
\begin{align}
x_n=\pi_n-\pi_{n+1}, \quad n=0,1,\ldots,N-1.
\end{align}
For completeness, we define \(x_N=\pi_N\). According to the monotonic constraints in \eqref{eq:lp_ex_1}, we have \(x_n\geq 0\), where \( n=0,1,\ldots,N\). Moreover, \(\pi_n\) can be recovered from \(x_n\) as
\begin{align}
\pi_n=\sum_{i=n}^{N}x_i. \label{eq:tf}
\end{align}
By substituting the above expression into the normalization constraint in \eqref{eq:lp_ex_2}, we obtain
\begin{align}
\sum_{n=0}^{N}\pi_n
=
\sum_{n=0}^{N}\sum_{i=n}^{N}x_i
=
\sum_{i=0}^{N}(i+1)x_i
=1.
\end{align}
Define
\begin{align}
y_i=(i+1)x_i,\quad i=0,1,\ldots,N.
\end{align}
Then, the constraints on \(y_i\) become
\begin{align}
y_i\geq0,\quad \sum_{i=0}^{N}y_i=1,
\end{align}
which correspond to a standard probability simplex. The extreme points of this simplex are given by
\begin{align}
y_n=1,\quad y_i=0,\quad i\neq n.
\end{align}
Accordingly, we have
\begin{align}
x_n=\frac{1}{n+1}, \quad x_i=0,\quad i\neq n.
\end{align}
By substituting these values into \eqref{eq:tf}, we obtain \eqref{eq:f_lp_ex_1} and \eqref{eq:f_lp_ex_2}, which completes the proof.
\end{IEEEproof}

\begin{figure}
  \centering
  \includegraphics[scale=0.67]{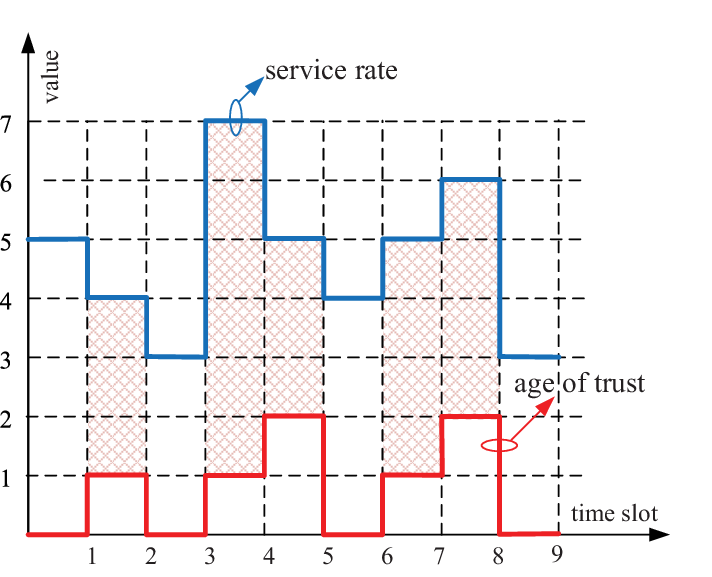}
  \caption{Continuous trust verification under a random service process, where $\alpha = 1$.}\label{fig:AoT_random}
\end{figure}

\subsubsection{Instantaneous Service Rate is Observable}
When the instantaneous service rate is known, we can further fine-tune the trust verification scheme. Specifically, given the verification decision $\mathcal{I}(t)$, the instantaneous reward is $\left[1-\mathcal{I}(t)\right]\mu(t) - \alpha\delta(t)$, which is jointly determined by the instantaneous service rate and the instantaneous AoT. An example of the random service process and the AoT process is shown in Fig.~\ref{fig:AoT_random}. We can see that the instantaneous reward, i.e., the height of the shadowing part at each slot, is randomly varying over time. It is difficult to use graphical methods to find the optimal verification scheme like the constant service process case. To tackle this problem, we consider using reinforcement learning, which is widely used to address various MDP problems over wireless networks~\cite{Cheng2022Adaptive,Zhou2026On}. In particular, $Q$-learning is one representative method of reinforcement learning, which includes four core elements, namely, environment space, action space, reward and value function. For more information of the basic principle of reinforcement $Q$-learning, please refer to the literatures \cite{van2016deep,Cui2023Multi}. We now come to how to map the MDP problem \textbf{P2} to the four core elements mentioned above.

\textbf{a) Environment Space:}
In accordance with \textbf{P2}, the environment state we can observe includes the instantaneous service rate at the current time slot and the AoT at the last time slot. So, the environment space, denoted by $\mathcal{S}$, is give by
\begin{align}
\mathcal{S} = \{s_t| s_t = (\mu(t),\delta(t - 1))\},
\end{align}
where $s_t$ is the environment state at time slot $t$.

\textbf{b) Action Space:}
We can select action between two candidates, that is, conducting the verification or performing the regular data transmission, corresponding to the afore defined indictor function $\mathcal{I}(t)$. Therefore, the action space, denoted by $\mathcal{A}$, is
\begin{align}
\mathcal{A} = \{a_t|a_t = \mathcal{I}(t)\},
\end{align}
where $a_t$ denotes the selected action at time slot $t$.

\textbf{c) Reward:}
Once an action is selected at the current environment state, a reward can be obtained to score it. For \textbf{P2}, the reward, denoted by $r(s_t,a_t)$, is given by
\begin{align}\label{eq:reward}
r(s_t,a_t) = \left[1-\mathcal{I}(t)\right]\mu(t) - \alpha\delta(t).
\end{align}

\textbf{d) Value Function:}
In $Q$-learning, the value function, denoted by $Q(s_t, a_t)$, indicates the long-term reward, which generally corresponds to the objective of MDP problems. Therefore, we have
\begin{align}
Q(s_t, a_t) = \sum_{t = 1}^{T}\left\{\left[1-\mathcal{I}(t)\right]\mu(t) - \alpha\delta(t)\right\}.
\end{align}
The value function also indicates the long-term gain of current action, updated by the state transition equation:
\begin{align}\label{eq:ste}
Q(s_t,a_t)= (1-\eta_{\rm lr})Q(s_t,a_t)+\eta_{\rm lr}[r_t+\gamma\max_a Q(s_{t+1},a)],
\end{align}
where $\eta_{\rm lr}$ is the learning rate and $\gamma$ is the discount factor. In this paper, the learning rate is set to 1. Moreover, to better approximate the average-reward objective, the discount factor is set close to one, which emphasizes long-term rewards~\cite{tsitsiklis2002average}.

\begin{algorithm}
\caption{Optimal Trust Verification Scheme via $Q$-Learning for Random Service Process}
\begin{algorithmic}[1]
\label{alg:opt_ver_scheme}
\STATE \textbf{input:} the service rate $\mu(1)$, the AoT $\delta(0)$, greedy probability $\varepsilon$, reduction period $t_\varepsilon$, and discount factor $\gamma$.
\STATE initialize time slot $t = 1$.
\STATE initialize environment state $s_t = (\mu(t),\delta(t - 1))$.
\WHILE {$Q(s_t,a_t)$ does not converge}
\STATE generate a random value $\varepsilon_t$ among 0 to 1.
\IF {$\varepsilon_t\leq\varepsilon$}
\STATE $a_t = 0\mbox{~or~}1$ with the equal probability.
\ELSE
\STATE $a_t = \arg\max Q(s_t,a_t)$.
\ENDIF
\STATE update AoT $\delta(t)$.
\STATE calculate the reward by using Eq.~(\ref{eq:reward}).
\STATE update $Q(s_t,a_t)$ by using Eq.~(\ref{eq:ste}).
\IF {$t\%t_\varepsilon == 0$}
\STATE reduce the value of $\varepsilon$.
\ENDIF
\STATE $t = t + 1$.
\STATE observe the service rate $\mu(t)$.
\STATE update the environment state $s_t = (\mu(t),\delta(t - 1))$.
\ENDWHILE
\STATE \textbf{output:} the optimal trust verification scheme.
\end{algorithmic}
\end{algorithm}

With the above mappings, we follow the procedure in Algorithm~\ref{alg:opt_ver_scheme} to find the optimal trust verification scheme. First, following the $\varepsilon$-greedy policy, we choose an arbitrary action or the action that maximizes $Q(s_t, a_t)$. Then, we update the value of AoT, the reward, and the value of the $Q$ function. The exploration probability is also updated, which follows a linear decay schedule, i.e., \(\varepsilon\) is reduced by a fixed amount every \(t_\varepsilon\) iterations until reaching zero. After that, we move to the next time slot and repeat the above process. Until $Q(s_t, a_t)$ converges, the optimal trust verification scheme is obtained. The convergence criterion is satisfied when either the maximum number of iterations is reached or the change in the moving-average reward between two consecutive evaluation windows is smaller than a predefined threshold.

Inspired by the periodic verification scheme for the random service process with the knowledge of the average service rate, we also propose an improved periodic verification scheme for the case of observable instantaneous service rate. This improved version dictates that verification should be performed not only at the beginning of the verification period, but also in the case of $\mu(t) - \alpha(\delta(t-1) + 1)\leq 0$. In addition, once verification is complete, a new verification cycle count will begin. Specifically, the verification decision is triggered when the immediate transmission reward is insufficient to compensate for the AoT penalty, i.e., $\mu(t)-\alpha(\delta(t-1)+1)\leq 0$. This condition represents a myopic reward-aware threshold, where the verification action is preferred when the expected loss caused by AoT increase outweighs the instantaneous transmission benefit. Note that this condition is jointly determined by the current service capability and the accumulated trust age, rather than merely indicating poor channel conditions.

It is worth noting that for the AoT minimization problems in this section, the standard optimization techniques, e.g., graphical methods and standard $Q$-learning, are used and proved to be effective. This solvability is a key advantage of our proposed AoT framework. It indicates that AoT captures the essential trust dynamics while maintaining mathematical tractability, making it a practical candidate for deployment in future zero trust networks where low-complexity algorithms are preferred.

\begin{figure}
  \centering
  \includegraphics[scale=0.75]{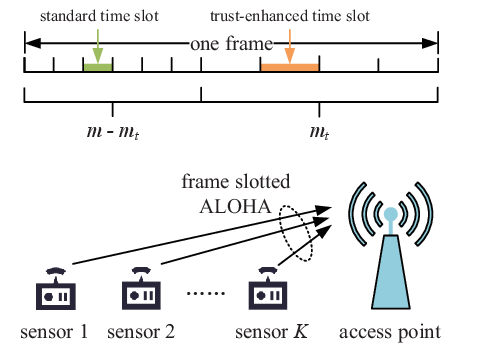}
  \caption{Continuous trust verification under the trust-enhanced frame slotted ALOHA.}\label{fig:T_FSA_access}
\end{figure}

\section{Continuous Trust Verification for Multiple Random Access}
\label{sec:AoT_app_two}
In this section, we discuss how to conduct the continuous trust verification for multiple random access and reach a fair compromise between AoT and throughput over a shared wireless channel. Firstly, the system model is elaborated on. Then, The optimal trust verification design is presented.
\subsection{System Model}
As shown in Fig.~\ref{fig:T_FSA_access}, we consider a scenario where $K$ sensors wish to deliver their status update information to the target access point over a shared wireless channel. Frame-slotted ALOHA is then used for multi-sensor access scheduling. Specifically, we denote by $T_f$ the frame period, which is further divided into $m$ time slots. During each frame period, each sensor is assumed to have a status packet to transmit with probability $\rho$ and remain idle with probability $1-\rho$, i.e., following a binomial distribution~\cite{Munari2021Modern}. In order to update its status packet to the access point in time, it will randomly occupy one time slot over a frame to load the packet. If no other sensor occupies this slot at the same time, the transmission will be successful. Otherwise, the transmission fails and the corresponding packet is discarded.

In contrast to traditional frame-slotted ALOHA, we slightly adjust the design of this protocol to perform continuous trust verification. Specifically, we design two types of time slots within a frame, that is, standard time slot and trust-enhanced time slot. As shown in Fig.~\ref{fig:T_FSA_access}, the standard time slot is only designed to transmit the status packet. However, the status packet could belong to the unauthorised sensors trying to access the wireless channel. To avoid this behaviour, we introduce the trust-enhanced time slot, which consists of a trust verification phase and a status packet transmission phase. The verification phase is used by the access point to extract a hardware fingerprint to identify the owner of the next status packet. We assume that the access point has collected all hardware fingerprints of authorised sensors at the beginning of the first access authorization of them and therefore the status packet generated by the unauthorised sensor will not be serviced.

In general, the length of the trust-enhanced time slot is greater than that of the standard time slot. It is a fact that increasing the percentage of trust-enhanced time slot will increase the security reliability, i.e. smaller AoT, but reduce the throughput of the channel, which is defined as how many status packets are successfully transmitted per unit time. In the following, we study how to fine-tune the percentage of trust-enhanced time slot to achieve the trade-off between AoT and throughput.

\begin{figure}
  \centering
  \includegraphics[scale=0.75]{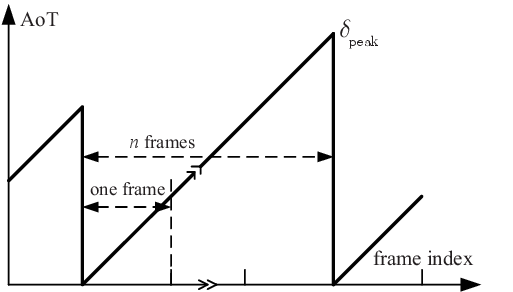}
  \caption{AoT process for each sensor.}\label{fig:AoT_access}
\end{figure}

\subsection{AoT Process and Throughput}
The successful transmission probability of status packet is essential for characterizing the AoT process and deriving the throughput for our focused scenario. Consider that a sensor is ready to transmit a fresh status packet to the access point, who randomly selects a time slot, either standard time slot or trust-enhanced time slot, to load its packet. Suppose that there are $\tilde{k}$ active sensors in the rest of $K-1$ sensors. Then, the successful transmission probability under this case is
\begin{align}
{\rm P}_{s,\tilde{k}} = \rho \left(1-\frac{1}{m}\right)^{\tilde{k}}.
\end{align}
The successful transmission probability by considering all possibilities is further given by~\cite{Feng2023Timely}
\begin{align}
{\rm P}_{s} = \sum_{\tilde{k} = 0}^{K - 1}\binom{K - 1}{\tilde{k}}\rho^{\tilde{k}}(1-\rho)^{K-1-\tilde{k}}{\rm P}_{s,\tilde{k}}
=\rho \left(1-\frac{\rho}{m}\right)^{K - 1}.
\end{align}
If the selected time slot is the trust-enhanced one, the identity of the sensor will be accordingly verified. We denoted by $m_t$ the number of trusted-enhanced time slots within one frame, and then the verification probability, denoted by ${\rm P}_{t}$, can be obtained as follows:
\begin{align}
{\rm P}_{t} = \frac{m_t}{m}{\rm P}_{s}.
\end{align}
Fig.~\ref{fig:AoT_access} shows the AoT process of each sensor, where AoT is linearly increasing as the frame index increases and reset to zero at the end of the frame in which the sensor's identify is successfully verified. Here, we absolutely trust the sensor if its identity is verified. Following this procedure, the distribution of the peak AoT, denoted by $\delta_{\rm peak}$, can be derived as follows:
\begin{align}
{\rm Pr}(\delta_{\rm peak} = n) = (1-{\rm P}_t)^{n-1}{\rm P}_{t}.
\end{align}
Based on the AoT process and the distribution of peak AoT, the average AoT, denoted by $\Delta$, can be obtained by
\begin{align}\nonumber
\Delta & = \frac{\displaystyle\sum_{n = 1}^{\infty}\frac{1}{2}n^2{\rm Pr}(\delta_{\rm peak} = n)}{\displaystyle\sum_{n = 1}^{\infty}n{\rm Pr}(\delta_{\rm peak} = n)} = \frac{\displaystyle\sum_{n = 1}^{\infty}n^2(1-{\rm P}_t)^{n-1}}{\displaystyle2\sum_{n = 1}^{\infty}n(1-{\rm P}_t)^{n-1}} \nonumber\\
& = \frac{2 - {\rm P}_{t}}{2 {\rm P}_{t}}. \label{eq:aver_aot_aloha}
\end{align}

Moreover, the throughput, denoted by $\eta$, following its definition is given by
\begin{align}\label{eq:throughput}
\eta = \frac{K{\rm P}_s}{T_f},
\end{align}
with
\begin{align}
T_f = m_t \beta T_s + (m - m_t)T_s,
\end{align}
where $T_s$ is the duration of one standard time slot and $\beta > 1$ is the ratio of the trust-enhanced time slot to the standard time slot.

\subsection{Optimal AoT-Aware Design for Frame Slotted ALOHA}
\begin{figure}
  \centering
  \includegraphics[scale = 0.6]{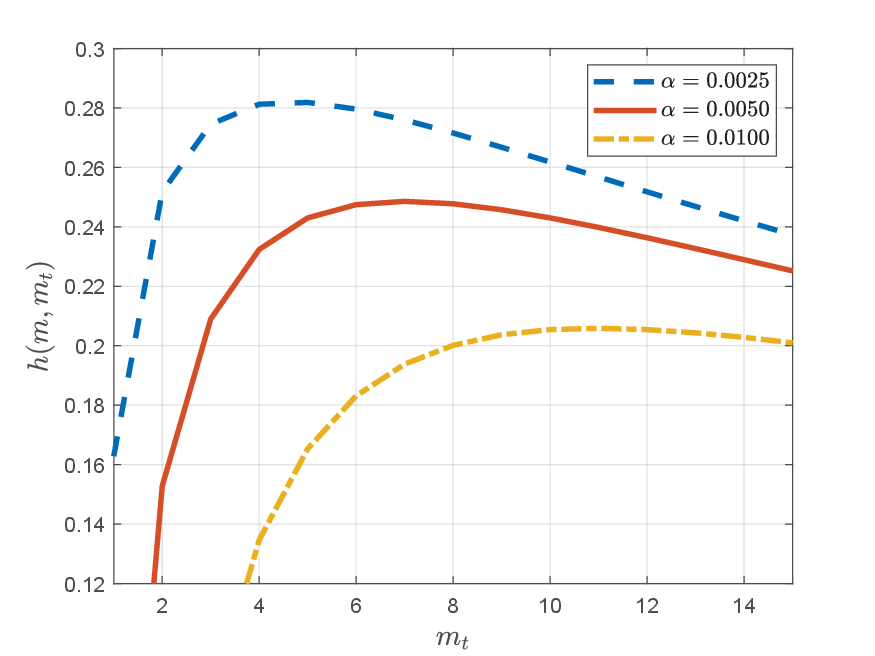}
  \caption{Analytical results of the objective function $h(m,m_t)$ versus $m_t$, where $K = 30$, $\rho = 0.5$, $\beta = 1.5$ and $m = 15$.}\label{fig:opt_m_t}
\end{figure}

\begin{figure}
  \centering
  \includegraphics[scale = 0.6]{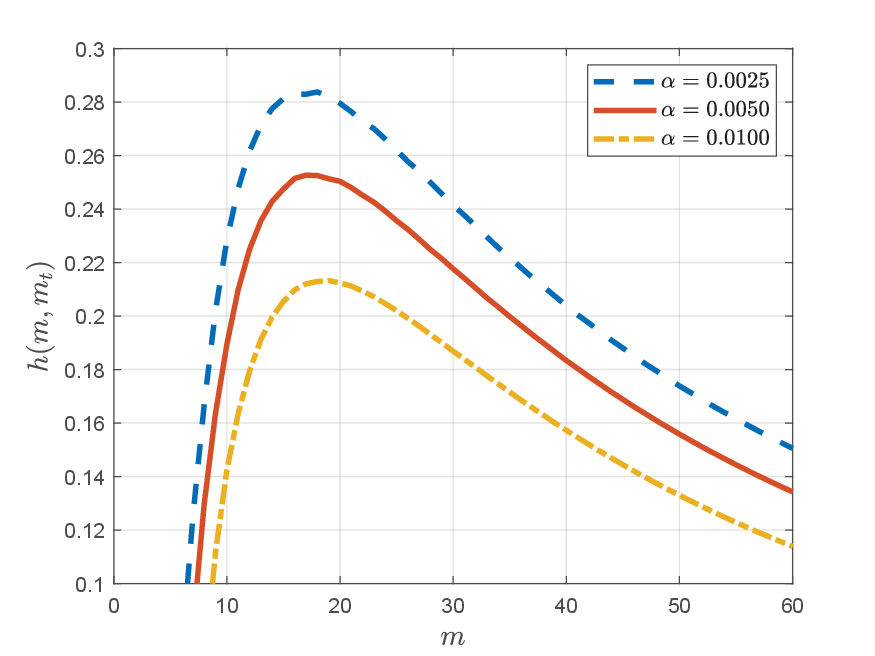}
  \caption{Analytical results of the objective function $h(m,m_t)$ versus $m$, where $K = 30$, $\rho = 0.5$, and $\beta = 1.5$.}\label{fig:opt_m}
\end{figure}

Our target is to obtain the numbers of standard and trust-enhanced time slots which can maximize the throughput meanwhile minimizing the average AoT. Mathematically, the corresponding optimization problem, denoted by \textbf{P3}, is formulated as follows:
\begin{align}
\textbf{P3}:
\max_{m_t,m} \,\,\, (\eta,-\Delta).
\end{align}
We use the sum weighted method to address this BOO problem. Then, \textbf{P3} is converted to \textbf{P4} as follows:
\begin{align}
\textbf{P4}:
\max_{m_t,m} \,\,\, \eta-\alpha\Delta,
\end{align}
where $\alpha$ is the weighted factor of average AoT. For ease of description, we use $h(m,m_t)$ to represent the objective of \textbf{P4}. By substituting Eqs.~(\ref{eq:aver_aot_aloha}) and (\ref{eq:throughput}) into $h(m,m_t)$, we have
\begin{align}\nonumber
h(m,m_t) &= \frac{K{\rm P}_s}{T_f}-\alpha\frac{2 - {\rm P}_{t}}{2 {\rm P}_{t}}\\
&= \frac{K{\rm P}_s}{m + (\beta - 1)m_t} - \alpha\left(\frac{2 - {\rm P}_{t}}{2 {\rm P}_{t}}\right),
\end{align}
where the duration of a standard time slot $T_s$ is assumed to be one without loss of generality~\cite{Munari2021Modern}. Fig.~\ref{fig:opt_m_t} illustrates the trend of $h(m,m_t)$ versus $m_t$, indicating that there exists the optimal $m_t$, and also, the optimal $m_t$ is increasing as the weighted factor of average AoT $\alpha$ increases. Motivated by this phenomena, we take the derivative of $h(m,m_t)$ with respect to $m_t$, i.e.,
\begin{align}
\frac{\partial h(m,m_t)}{\partial m_t}= \frac{\alpha m}{{\rm P}_sm_t^2} - \frac{(\beta - 1)K{\rm P}_s}{(m + (\beta - 1)m_t)^2},
\end{align}
and then setting the result to be zero, we obtain the optimal $m_t$ for a given $m$ as follows:
\begin{align}\label{eq:opt_m}
m_t = \frac{m\sqrt{\alpha m}}{\sqrt{K(\beta - 1)}{\rm P}_s - \sqrt{\alpha m}(\beta - 1)}.
\end{align}
Further, since $m_t$ should be an integer, we ceil or floor the above result, and then the one yielding a higher objective value of Problem~\textbf{P4} is selected. Substituting the optimal $m_t$ into $h(m,m_t)$, $h(m,m_t)$ becomes a single-variable yet complex function with respect to $m$. Since $m$ is an integer, we consider using the exhaustive method to search the optimal $m$. Fig.~\ref{fig:opt_m} shows the trend of the updated $h(m,m_t)$ with respect to $m$. It can be seen that there exists the optimal $m$, and it does not affected by the variation of the weighted factor of average AoT.

\emph{Remark}: It is important to highlight that the optimization problem P3 benefits significantly from the closed-form derivation of the average AoT in (\ref{eq:aver_aot_aloha}). While the subsequent optimization leverages standard derivation, this analytical tractability is a deliberate and valuable feature for MAC design. Unlike complex numerical methods often required in non-convex network optimization, the derived closed-form solution allows for the direct calculation of the optimal frame configuration, as shown in Eq. (\ref{eq:opt_m}). This is particularly critical for IoT sensors with limited processing capabilities.

Now we have presented the derivation of the optimal numbers of standard time slots and trusted-enhanced time slots associated with the specified weighted factor of average AoT $\alpha$. In practical applications, we can adjust $\alpha$ catering to various systems' security requirements and reach the optimal compromise between throughput and AoT.

\section{Simulation Results}
\label{sec:numerical_results}
In this section, we verify the effectiveness of our proposed schemes. First, we evaluate the performances of our proposed schemes for continuous trust verification in single-link transmissions. Then, we evaluate the performances of the trust-enhanced frame-slotted ALOHA for continuous trust verification in a multiple random access scenario. The used simulation software is Matlab R2022a. Unless stated otherwise, the simulation parameters are configured as follows. For the single-link transmission scenario, the number of iterations is set to $3\times 10^4$ and the weighted factor of the average AoT is set to $\alpha = 1$. For the constant service process, the service rate is set to $\mu(t) = 7$ packets/slot. For the random service process, we consider a binomially distributed service process, where the service rate is set to $\mu(t) = 1$ or $10$ packets/slot with the equal probability. We also consider a random service process over a Rayleigh wireless fading channel, where the size of a packet is set to \(L = 1000\) bits, the bandwidth is set to \(B = 1\) MHz, the duration of a time slot is set to \(T_s = 10\) ms, and the average power gain is set to $\bar{\gamma}= 0$ dB. At each time slot, the instantaneous achievable channel transmission rate \(R(t)\) is first calculated, and then \(\mu(t)\) is obtained as an integer by quantizing the transmitted bits according to the packet size, i.e., \(\mu(t)=\lfloor\frac{R(t)T_s}{L}\rfloor\). Since $\mu(t)$ is discretized as an integer and the maximum value of it is generally small, a finite and discrete $Q$-table can be constructed. Moreover, a randomized verification scheme is used as the baseline, where the trust verification is performed with a probability of $\phi$ at each time slot and the value of $\phi$ is varied from 0 to 1 to reach a compromise between the average AoT and throughput. For the multiple random access scenario, the simulation results are averaged over $10^6$ time slots and the weighted factor of the average AoT is set to $\alpha = 0.1$. Furthermore, the transmission probability is set to $\rho=0.5$. The number of time slots in one frame is set to $m = 12$. The number of sensors is set to $K=20$. The verification overhead ratio is set to $\beta = 1.5$.

\vspace{-10pt}
\subsection{Performance Evaluation for Single-Link Transmissions}
\begin{figure}
  \centering
  \includegraphics[scale = 0.6]{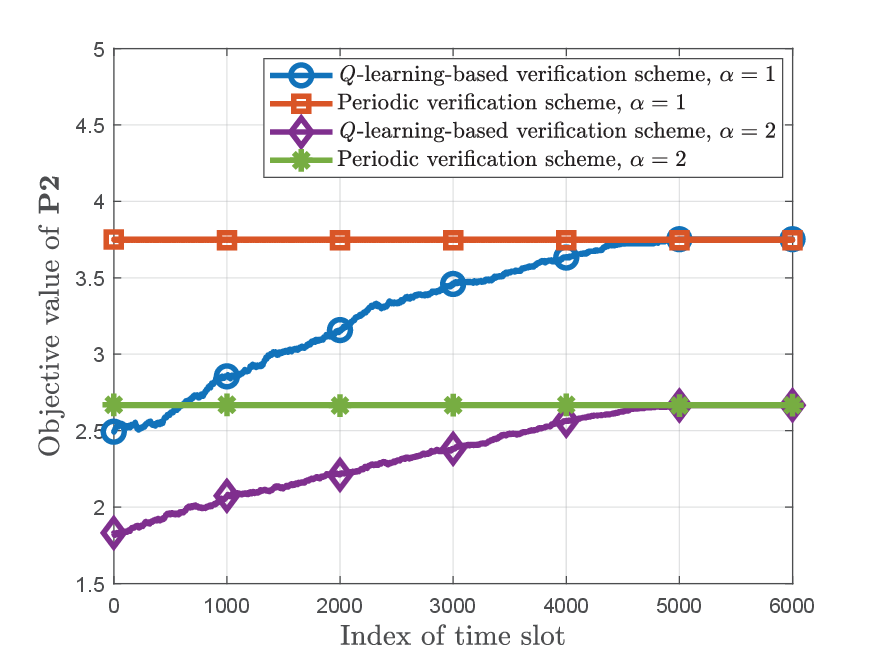}
  \caption{Simulation result showing the performances of various verification schemes under the given constant service process.}\label{fig:performance_p2_constant}
\end{figure}

\begin{figure}
  \centering
  \includegraphics[scale = 0.6]{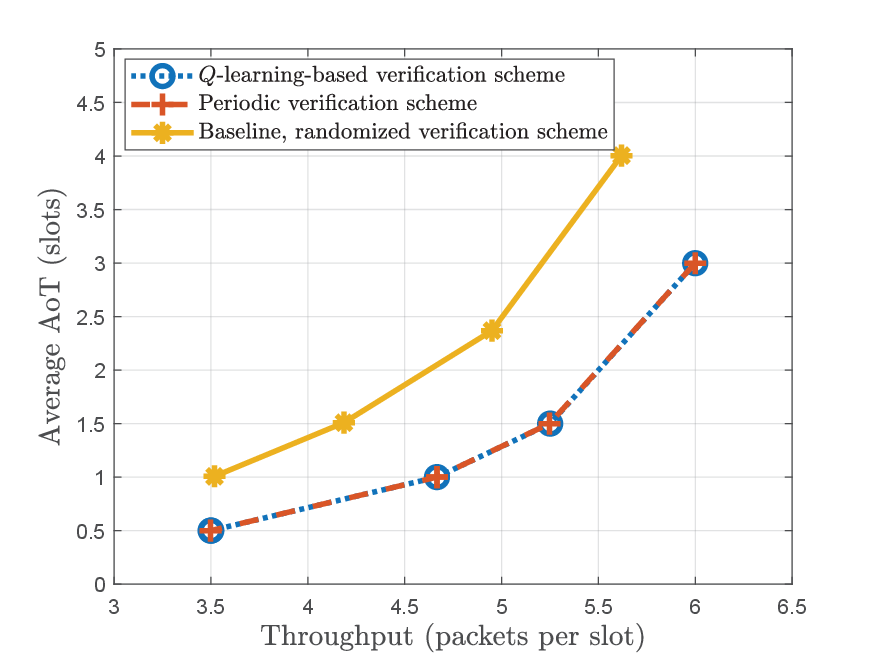}
  \caption{Simulation result showing the trade-offs between average AoT and throughput with various verification schemes under the given constant service process.}\label{fig:compromise_constant}
\end{figure}

In Fig. 9, we validate the convergence and optimality of the proposed $Q$-learning-based verification scheme under the given constant service process. It can be clearly observed that the $Q$-learning scheme starts with a lower objective value due to the initial exploration phase, i.e., the $\epsilon$-greedy policy. However, as the number of iterations increases, the performance curve gradually climbs and eventually converges to the exact same level as the optimal periodic verification scheme derived in Section III-B. Specifically, the convergence is achieved around the 4,500th time slot. This phenomenon confirms that the reinforcement learning agent successfully learns the optimal verification period without prior knowledge of the environments. Furthermore, by comparing the curves for different weighting factors, we see that the objective value decreases as $\alpha$ increases from 1 to 2. The reason is that a larger $\alpha$ imposes a stricter penalty on the AoT, forcing the system to perform verification more frequently. This increased verification overhead consumes time slots that would be used for data transmission, thereby reducing the overall system throughput and the objective value.

\begin{figure}
  \centering
  \includegraphics[scale = 0.6]{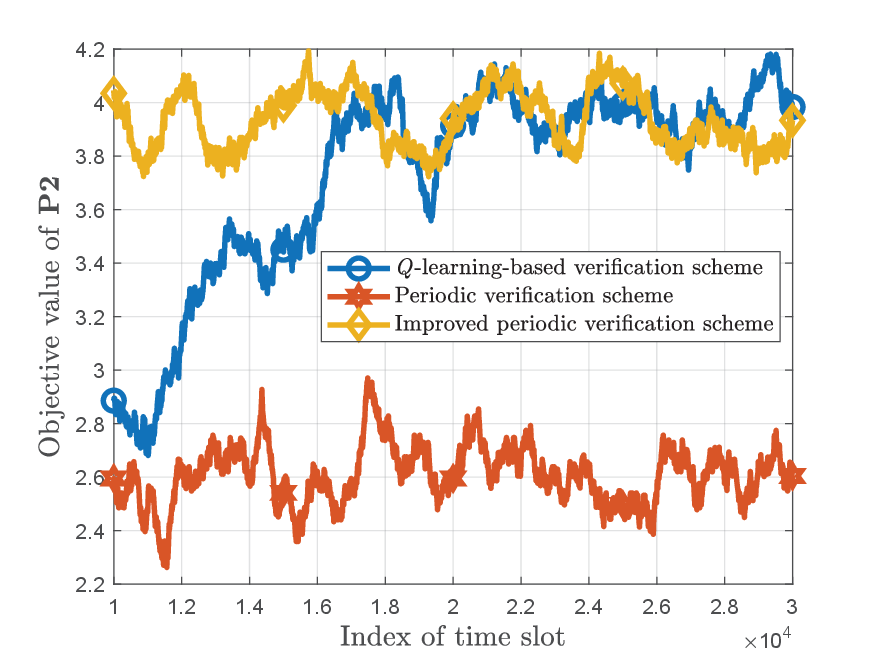}
  \caption{Simulation result showing the performances of various verification schemes under the given binomially distributed service process.}\label{fig:performance_p2_random}
\end{figure}

\begin{figure}
  \centering
  \includegraphics[scale = 0.6]{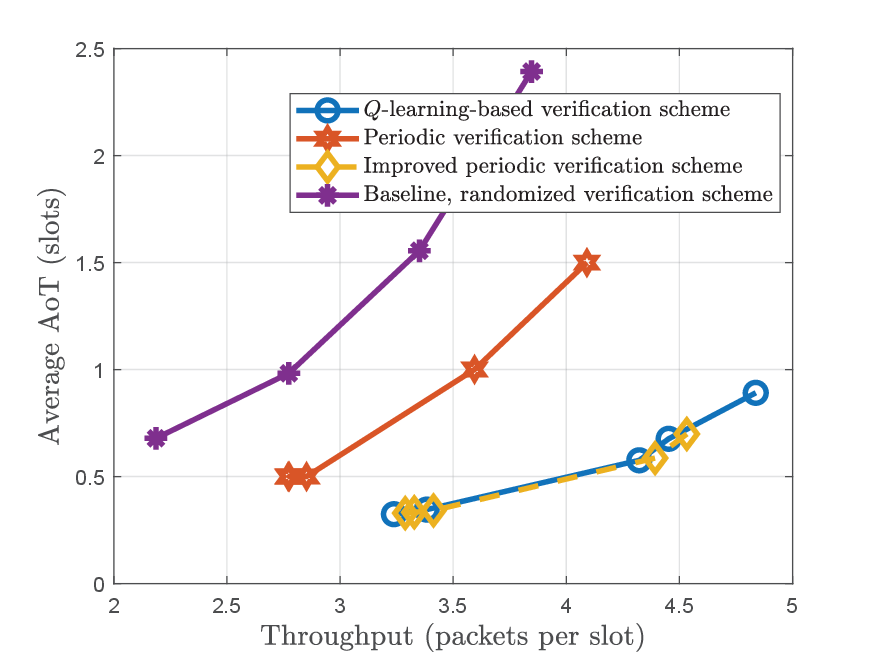}
  \caption{Simulation result showing the trade-offs between average AoT and throughput with various verification schemes under the given binomially distributed service process.}\label{fig:compromise_random}
\end{figure}

Fig.~\ref{fig:compromise_constant} shows the trade-offs between average AoT and throughput with our proposed verification schemes and the baseline scheme under the given constant service process. It has been shown in the previous figure that the same objective of \textbf{P2} can be achieved by using our proposed $Q$-learning based scheme and periodic verification scheme. This leads to the same compromise between average AoT and throughput in Fig.~\ref{fig:compromise_constant} with the use of these two schemes. Moreover, we can observe that the proposed $Q$-learning-based and periodic verification schemes achieve a better average AoT-throughput trade-off compared to the baseline scheme. Specifically, to achieve an average AoT of 1.5 slots, our scheme maintains a throughput of 5.3 packets/slot, whereas the baseline scheme drops to 4.2 packets/slot. The reason lies in the fact that the $Q$-learning-based and periodic verification schemes perform verification in a timely manner. Also, as the throughput decreases, the average AoT decreases first rapidly and then slowly, which indicates that we can sacrifice some throughput performance in a reasonable range to significantly reduce the average AoT, and sacrificing too much throughput performance may not be cost effective.

Fig.~\ref{fig:performance_p2_random} highlights the importance of exploiting instantaneous service rate in random service processes. It demonstrates that schemes capable of observing the instantaneous service rate, i.e., the $Q$-learning-based and improved periodic schemes, significantly outperform the strictly periodic verification scheme that relies solely on average statistics. Specifically, we evaluate the performance of the proposed schemes under the given binomially distributed service process. We can observe that the improved periodic verification scheme maintains a high and stable objective value throughout the process. In contrast, the $Q$-learning-based scheme starts with a lower performance due to its exploration-exploitation trade-off but rapidly converges to the same high-performance level as the improved periodic scheme after approximately 2,0000 time slots. Notably, both schemes achieve an objective value of around 4.0, representing a substantial gain over the strictly periodic verification scheme, which saturates at roughly 2.6. The underlying reason for this performance gap lies in the opportunistic nature of the superior schemes. As discussed in Section III-C, the $Q$-learning-based and improved periodic schemes can identify the time slots with low instantaneous service rates to perform verification. This strategy effectively resets the AoT with minimal throughput penalty. Conversely, the strictly periodic verification scheme, being oblivious to instantaneous variations, fails to seize these low-cost opportunities, leading to suboptimal verification decision.

\begin{figure}
  \centering
  \includegraphics[scale = 0.6]{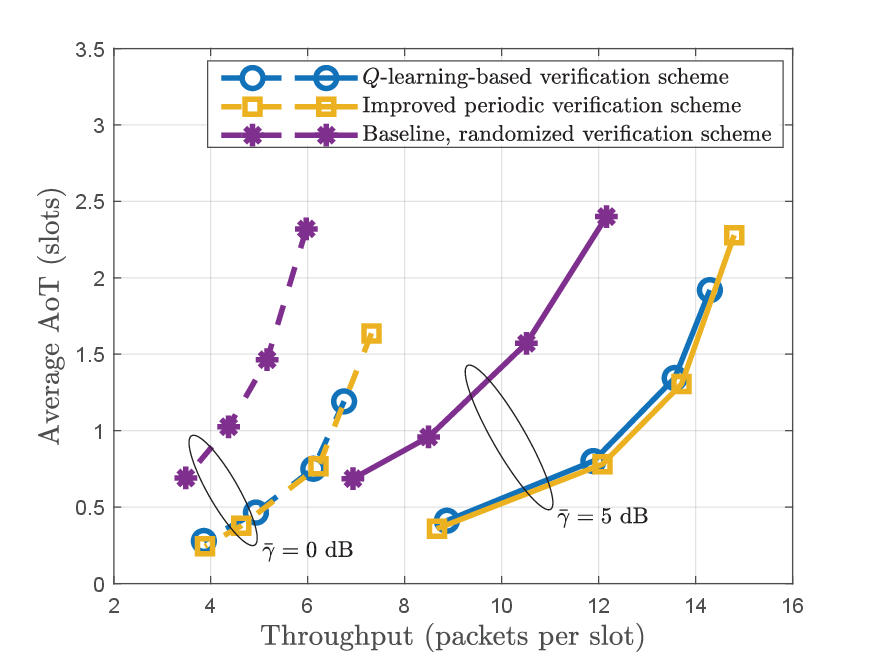}
  \caption{Simulation result showing the trade-offs between average AoT and throughput with various verification schemes under the random service process over a Rayleigh wireless fading channel.}\label{fig:compromise_random_Ray}
\end{figure}

\begin{figure}
  \centering
  \includegraphics[scale = 0.6]{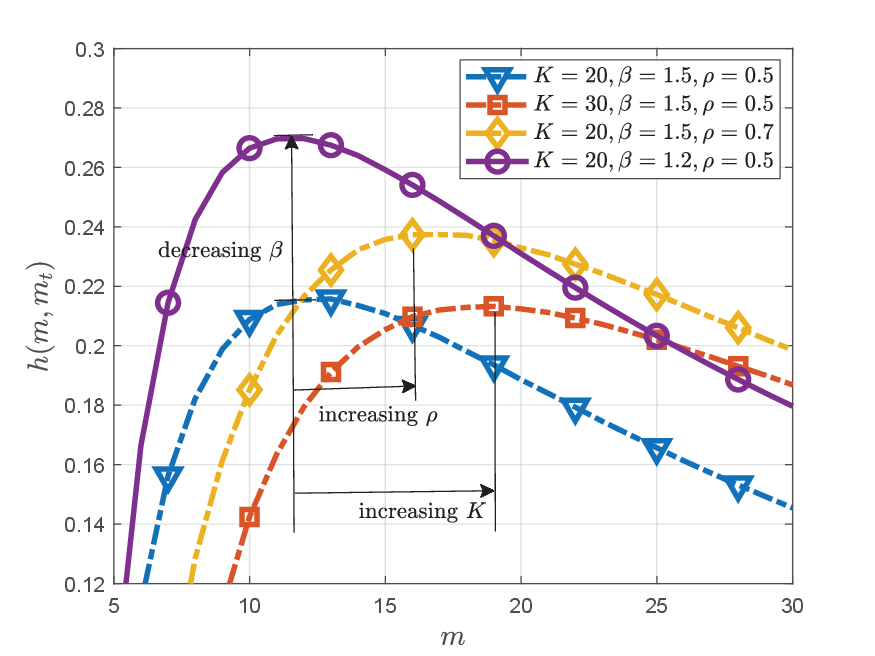}
  \caption{Analytical results of the objective function $h(m,m_t)$ versus $m$, where $\alpha = 0.01$.}\label{fig:opt_m_analysis}
  \vspace{-10pt}
\end{figure}

In Fig.~\ref{fig:compromise_random}, we illustrate the trade-offs between average AoT and throughput with our proposed schemes and the baseline scheme under the given binomially distributed service process. In particular, it shows the superiority of the proposed $Q$-learning-based and improved periodic schemes, which achieve a better average AoT-throughput trade-off compared to the periodic verification scheme and the baseline scheme. As observed from the curves, for a given requirement of average AoT, e.g., 1 slot, the $Q$-learning-based and improved periodic schemes can support a throughput of approximately 4.8 packets/slot, whereas the periodic verification scheme is limited to roughly 3.6 packets/slot. This represents a throughput gain of approximately 33\% for the same security level. Conversely, the baseline scheme performs the worst, suffering from significantly lower throughput for any given AoT level. This is expected since the baseline scheme exploits neither the instantaneous nor the average service rate.

Fig.~\ref{fig:compromise_random_Ray} validates the robustness and practical applicability of the proposed schemes under a realistic Rayleigh fading channel model. It demonstrates that the proposed $Q$-learning-based and improved periodic verification schemes maintain their superiorities even when the service rate fluctuates continuously due to channel fading. Specifically, the figure plots the trade-off curves under different average channel power gains, denoted by $\bar{\gamma}$. We can observe that as $\bar{\gamma}$ increases from 0 dB to 5 dB, the average AoT-throughput trade-off curves of all schemes move toward a more favorable region, i.e., achieving higher throughput and lower AoT. Notably, under the same channel condition, the proposed $Q$-learning-based and improved periodic schemes both achieve a lower average AoT compared to the baseline scheme for any given throughput target. The underlying reason for these results is the adaptive capability of our proposal in exploiting channel diversity. In a Rayleigh fading environment, the channel state varies continuously. The $Q$-learning-based and improved periodic schemes perform trust verification during deep fade instances. This strategy minimizes the verification cost, thereby efficiently resetting the AoT without significantly penalizing the overall throughput. Furthermore, a higher $\bar{\gamma}$ naturally provides a larger resource budget, alleviating the contention between verification and transmission.

\begin{figure}
  \centering
  \includegraphics[scale = 0.6]{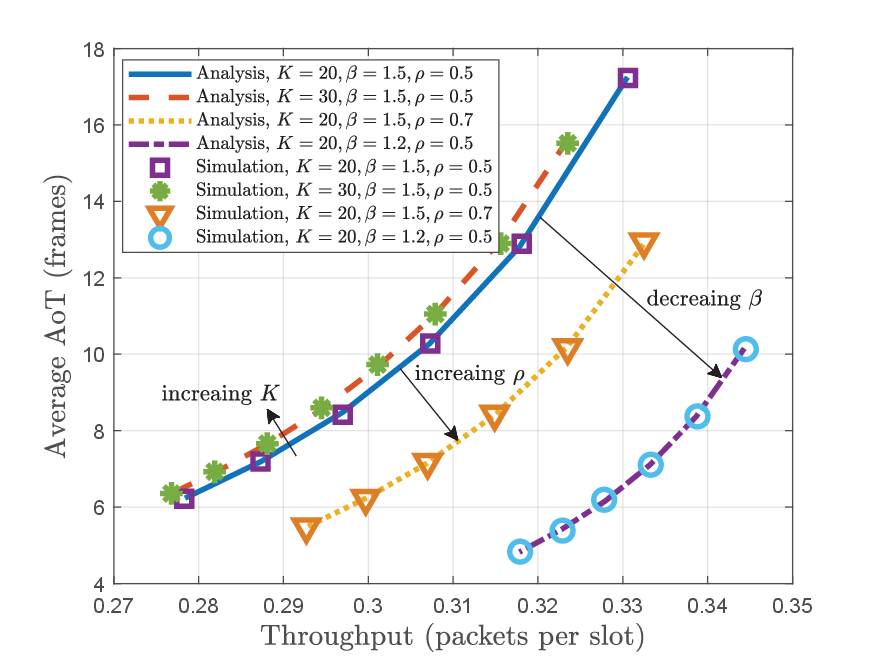}
  \caption{The trade-offs between average AoT and throughput with different numbers of sensors, transmission probabilities and verification overhead ratios.}\label{fig:compromise_multiple_access}
\end{figure}

\begin{figure}
  \centering
  \includegraphics[scale = 0.6]{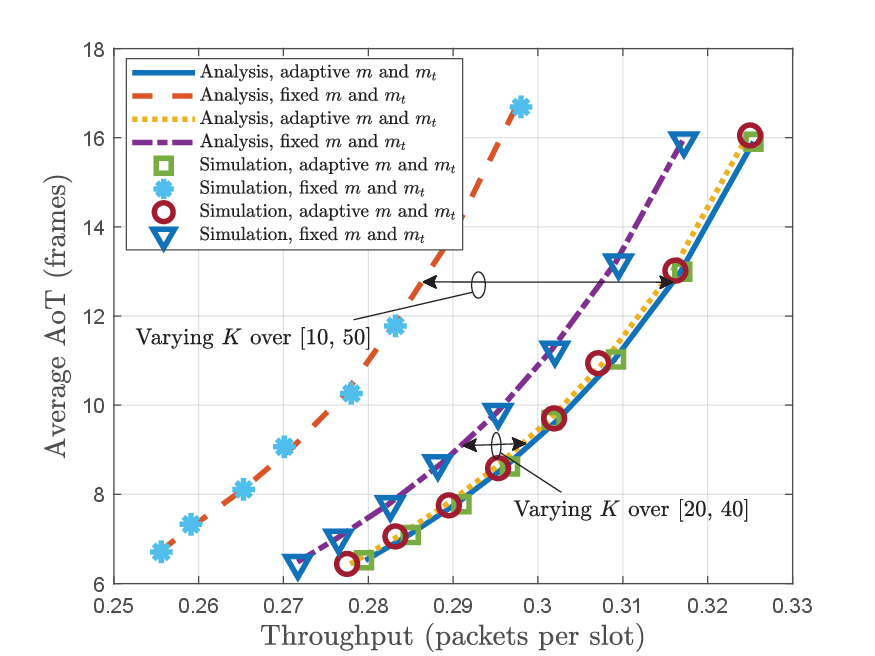}
  \caption{The trade-offs between average AoT and throughput under time-varying numbers of sensors.}\label{fig:compromise_multiple_access_DFSA}
\end{figure}

\subsection{Performance Evaluation for Multiple Random Access}

Fig.~\ref{fig:opt_m_analysis} shows the sensitivity of the optimal frame design to network traffic characteristics and its robustness to verification overhead. Specifically, by comparing the curves with different parameter settings, three key observations can be made. First, as the number of sensors $K$ increases, the optimal frame length $m$ increases. Second, a similar trend is observed for the transmission probability $\rho$ that increasing $\rho$ from 0.5 to 0.7 also necessitates a larger optimal $m$. Third, and most interestingly, decreasing the verification overhead ratio $\beta$ has a negligible impact on the optimal frame length itself. The underlying reasons for these phenomena are rooted in the contention mechanism of random access. For the first two observations, increasing either the user density $K$ or the transmission probability $\rho$ intensifies the traffic load, thereby enlarging the collision probability. Consequently, a larger frame length $m$ is required to alleviate contention to maintain a high successful transmission probability. Regarding the third observation, while $\beta$ affects the duration of the frame, the optimal number of slots $m$ is dominantly determined by the traffic load, i.e., the ratio of active users to available slots, rather than the duration of individual slots. Therefore, the optimal configuration of the frame length remains robust against variations in the verification time cost.

In Fig.~\ref{fig:compromise_multiple_access}, we investigate how different parameters affect the system performance. It is observed that, under the considered parameter settings, increasing the number of sensors \(K\) degrades the system performance, while increasing the transmission probability \(\rho\) improves the average AoT-throughput trade-off by shifting the curve toward a more favorable region. This observation highlights that the impact of traffic parameters in contention-based protocols depends on the balance between collision risk and verification opportunities. Specifically, three key observations can be drawn from the simulation and analytical results. First, as the number of sensors \(K\) increases, the average AoT-throughput trade-off curve shifts to the left, indicating degradation in both throughput and average AoT. Second, within the considered range of \(\rho\), increasing the transmission probability \(\rho\) shifts the curve to the right, since more transmission attempts provide additional opportunities for successful verification and AoT reduction. Third, decreasing the verification overhead ratio \(\beta\) naturally expands the achievable region. Additionally, the alignment between the simulation markers and the analytical curves validates the accuracy of our theoretical derivation in Section IV. The different impacts of \(K\) and \(\rho\) originate from their distinct effects on collision risk and verification opportunities. While increasing \(K\) mainly introduces additional contention without improving individual verification opportunities, increasing \(\rho\) has the potential to enable sensors to more actively access verification opportunities at the cost of higher collision probability. However, an excessively large \(\rho\) may increase packet collisions in frame-slotted ALOHA, and the optimal verification ratio depends on the specific system parameters.

Fig.~\ref{fig:compromise_multiple_access_DFSA} validates the effectiveness of our obtained frame configuration for dynamic frame slotted ALOHA~\cite{Schoute1983Dynamic}, which is more efficient when the number of users $K$ is time-varying and widely deployed in various machine communication networks~\cite{Moradian2024Age,George2018Performance}. It demonstrates that by dynamically adapting the frame configuration $m$ and $m_t$ according to the varying number of users $K$, the system achieves significantly higher performance compared to static configurations. Specifically, the figure compares the performance of the adaptive scheme, where $m$ and $m_t$ are optimized for each instantaneous $K$, against the fixed scheme, where parameters are fixed based on the average $K$. In the simulation, the average $K$ is 30. We can observe that the adaptive scheme consistently achieves a better average AoT-throughput trade-off, yielding lower average AoT for the same throughput compared to the fixed scheme. Moreover, the performance gain widens as the population of users increases. As shown in the figure, the gap between the adaptive and fixed schemes under the range $K \in [10, 50]$ is significantly larger than that under the narrower range $K \in [20, 40]$. The underlying reason for these results is the mismatch penalty inherent in static designs. In the fixed scheme, the frame parameters are only optimal for a specific average load. When the instantaneous user count $K$ deviates from this average, the fixed frame becomes suboptimal. In contrast, the adaptive scheme effectively tracks the optimal point. By leveraging the derived relationship between $K$ and the optimal $m$ and $m_t$, it eliminates the mismatch penalty, ensuring that the trust verification is performed efficiently regardless of traffic fluctuations.

\vspace{-10pt}
\section{Conclusions}
\label{sec:conclusions}
We proposed the concept of AoT, which is defined as the elapsed time since the target user's trust was last verified. This concept is expected to capture the characteristics of declining trust levels over time and to evaluate the level of continuous trust verification. Furthermore, since continuous trust verification requires a certain level of resources, we showed how to reach a fair compromise between AoT and other network performances. Specifically, we designed the optimal trust verification schemes in single-link transmissions. We also designed a trust-enhanced frame-slotted ALOHA for continuous trust verification in a multiple random access scenario. Numerical results showed that our proposals can achieve a fair compromise between the level of continuous verification and other network performances, and have great potential for applications in various zero-trust architectures.

In future work, we would extend our work to the third-party verification architecture. In such distributed scenarios, the AoT reset value will be non-zero, determined by the round-trip latency between the receiver and the remote third-party verifier. Investigating the impact of varying backhaul latencies on the AoT scheduling policy remains an open and interesting topic. We would also like to extend the single-link optimization in Section III to multi-user scenarios with heterogeneous security requirements. For instance, scheduling resources between a security-sensitive user (minimizing AoT) and a throughput-oriented user (maximizing rate) presents a fundamentally different challenge involving inter-user resource competition and game-theoretic fairness, which differs from the intra-user trade-off addressed in this work to find a Pareto optimal point tailored to one user's preference.

\bibliographystyle{IEEEtran}
\bibliography{References}

\end{document}